\documentstyle[aps,prl,multicol,epsfig]{revtex}
\begin{document}
\draft
\preprint{}
\wideabs{
\title{Effect of Spin Gap on Single-Hole Excitation Spectrum in the One-Dimensional $t$-$J$-$J'$ Model}
\author{Takami Tohyama and Sadamichi Maekawa}
\address{Institute for Materials Research, Tohoku University, Sendai 980-77, JAPAN}
\maketitle
\begin{abstract}

We examine the effect of spin gap on single-hole spectral function in the one-dimensional $t$-$J$-$J'$ model.  At $J'$/$J$=1/2, where a dimer state is the ground state, an analytic expression is derived for $J$/$t$$\rightarrow$0. At finite values of $J$/$t$, the numerical diagonalizations for finite-size clusters are performed.  In contrast to the gapless case, we find that (i) the dispersion becomes symmetric around $k$=$\pi$/2, (ii) the singularity of the holon band is strongly suppressed, and (iii) the quasi-particle weight $Z_{\rm h}$ becomes finite at finite $J/t$.  The feature (iii) is related to the presence of a bound state of the doped hole and an unpaired spin.
\end{abstract}}
\narrowtext
Recently, angle-resolved photoemission (ARPES) experiment on one-dimensional (1D) insulating copper oxide, SrCuO$_2$, has shown asymmetric dispersions around $k$=$\pi$/2~\cite{Kim}.  Between $k$=0 and $\pi$/2, two bands with the width of 0.3 and 1.2~eV have been observed, while a single 1.2~eV energy dispersion has been obtained between $k$=$\pi$/2 and $\pi$.  From the comparison with the exact diagonalization results of the single-hole spectral function for the 1D $t$-$J$ model, the two bands having 0.3 and 1.2~eV width have been identified as spinon and  holon  bands, respectively.

SrCuO$_2$ is a typical $S$=1/2 Heisenberg system with nearest-neighbor antiferromagnetic (AF) interaction $J$.  The spin excitation is thus gapless.  On the other hand, 1D compounds such as CuGeO$_3$ show a spin gap.  How does the spin gap affect on the spectral function?  How do the spinon and holon bands look like under the spin gap?  In order to answer these questions, we investigate the single-hole spectral function for spin-gapped 1D models.

The simplest 1D model with spin gap is the $S$=1/2 $J$-$J'$ model where $J'$ is next-nearest-neighbor AF interaction.  The minimum value of $J'$/$J$ to produce the spin gap is estimated to be 0.241~\cite{Okamoto}.  At $J'$/$J$=1/2 the ground state is a simple dimer state as proven by Majumdar and Ghosh~\cite{Majumdar}.
The Hamiltonian of the $t$-$J$-$J'$ model, $H=H_t +H_{JJ'}$, is written as
\begin{eqnarray}
H_t&=&-t\sum\limits_i(\tilde{c}_{i\sigma }^\dagger \tilde{c}_{j\sigma } + h.c.)\\
H_{JJ'}&=&J\sum\limits_i({\bf S}_i\cdot {\bf S}_{i+1}-{1 \over 4}n_in_{i+1})\nonumber\\
&+&J'\sum\limits_i({\bf S}_i\cdot {\bf S}_{i+2}-{1 \over 4}n_in_{i+2})\;,
\label{tJJ'}
\end{eqnarray}
where $\tilde{c}_{i\sigma}=c_{i\sigma}(1-n_{i\bar{\sigma}})$ is the annihilation operator of an electron with spin $\sigma$ at site $i$ with the constraint of no double occupancy, ${\bf S}_i$ is the spin operator, and $n_{i\sigma}=c_{i\sigma}^\dagger c_{i\sigma}$,  $n_i=n_{i\uparrow}+n_{i\downarrow}$.

The spectral function for up-spin electron removal is expressed as
\begin{equation}
A(k,\omega)=\sum\limits_{\nu}
\left|\left< \psi_{L,\nu}^h \left| c_{k\uparrow} \right|\psi_L^0 \right>\right|^2
\delta(\omega+ E_\nu^h - E_0)\;,
\label{Akw}
\end{equation}
where $\left|\psi_L^0\right\rangle$ denotes the ground state of the $J$-$J'$ model with energy $E_0$, and $\left|\psi_{L,\nu}^h\right\rangle$ denotes the $\nu$-th eigenvector of the one-hole $t$-$J$-$J'$ model with eigenvalue $E_\nu$.  $L$ is the total number of sites for a periodic ring.

In the case of $J'$/$J$=0 and $J$/$t$$\rightarrow$0, the wavefunction is expressed as a product of the wavefunction for spinless fermions and that for a squeezed Heisenberg model~\cite{Ogata1}.  The single-hole spectral function is then expressed as~\cite{Sorella}
\begin{equation}
A(k,\omega)=\sum\limits_{Q} Z(Q)\delta(\omega- E_{k,Q})\;,
\label{Akw0}
\end{equation}
where
\begin{eqnarray}
Z(Q)&=&{1\over 2}{1\over L-1}\sum\limits_{j=0}^{L-2} (-1)^je^{-iQj}\left<\psi_L^0\left| T_{(0,j)} \right|\psi_L^0\right>\;,
\label{ZQ}\\
T_{(0,j)}&=&(2{\bf S}_j\cdot{\bf S}_{j-1}+1/2)\ldots(2{\bf S}_1\cdot{\bf S}_0+1/2)
\label{T}
\end{eqnarray}
and
\begin{equation}
E_{k,Q}=-2t \cos (k-Q)\;.
\label{EkQ}
\end{equation}
$Q$ is the momentum coming from spin degree of freedom.  Note that $Z(Q)$ determines the weight of the spectral function.

We assume that, even for $J'$/$J$=1/2, the ground-state wavefunction can be expressed as the product of the two components~\cite{Ogata2}.  At $J'$/$J$=1/2, the ground-state wavefunction is exactly expressed as~\cite{Majumdar}
\begin{equation}
\psi_L^0=(\Phi_1+\Phi_2)/\sqrt{2\left(1+(1/2)^{L/2-1}\right)}\;,
\label{Dimer}
\end{equation}
for the total momentum of zero~\cite{Momentum} and $L$=4$n$+2 ($n$ being integer), where
\begin{eqnarray}
\Phi_1&=&[0,1][2,3]\ldots[L-2,L-1]\;,\\
\Phi_2&=&[1,2][3,4]\ldots[L-1,1]\;,
\end{eqnarray}
using singlet bond $[l,m]=(\uparrow_l\downarrow_m-\downarrow_l\uparrow_m)$.  It is straightforward to obtain $Z(Q)$ for large $L$:
\begin{eqnarray}
&&(L-1)Z(Q)=\nonumber\\
&&{1\over 2}\left[\cos Q-1+{2-\cos Q+\cos(2Q)-(1/2)\cos(3Q) \over 5/4+\cos(2Q)}\right]\;.
\end{eqnarray}
$Z(Q)$ is plotted in Fig.~1(a).  For comparison, diagonalization results and a fitted curve for the Heisenberg model ($J'$/$J$=0)~\cite{Sorella,Penc} are also shown.  There is a divergence at $Q$=$\pi$/2 for $J'$/$J$=0.  On the contrary, the divergent behavior is completely suppressed for $J'$/$J$=1/2 and sizable weight appears above $Q$=$\pi$/2.

\begin{figure}[t]
\begin{center}
\epsfig{file=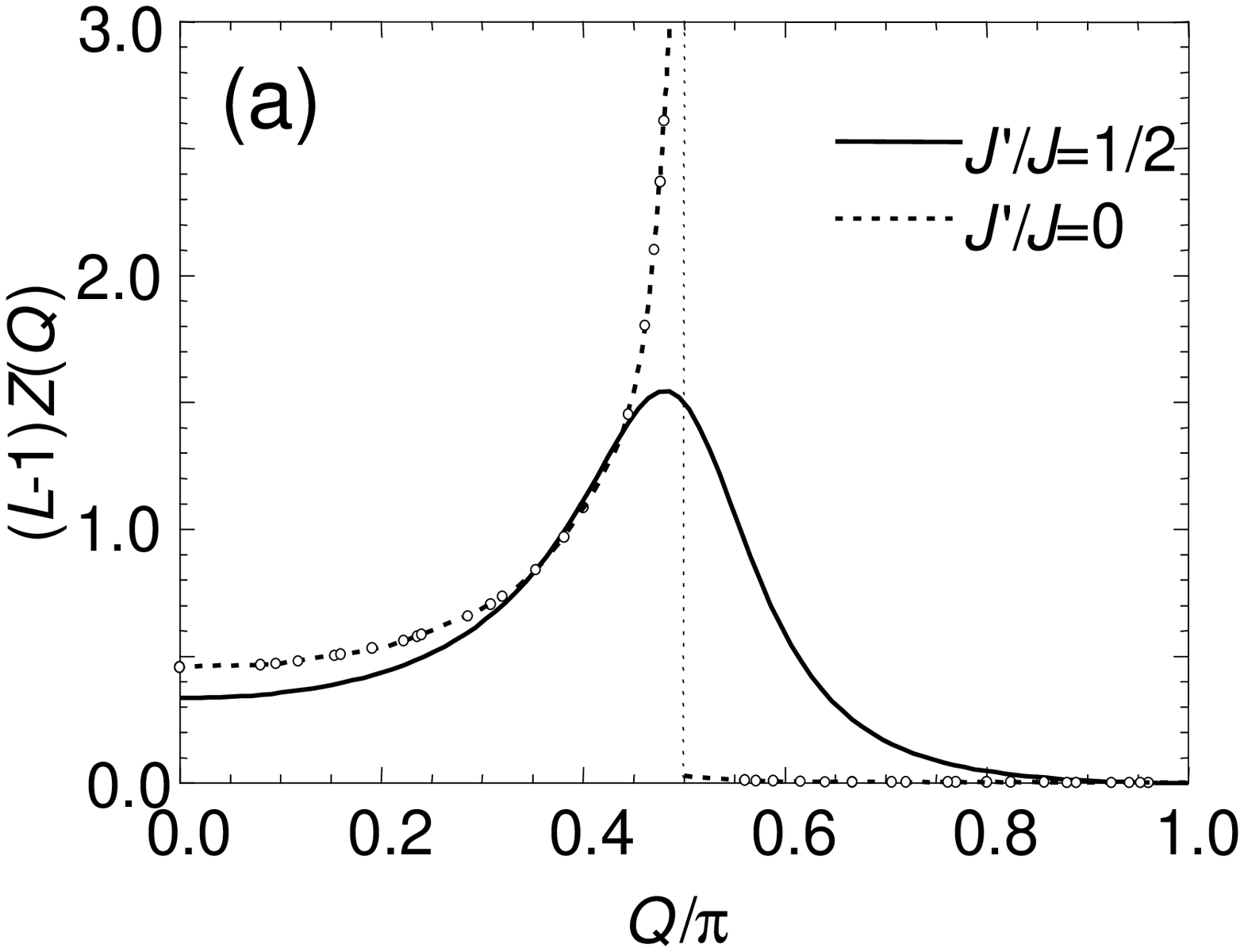,width=8cm,clip=}
\epsfig{file=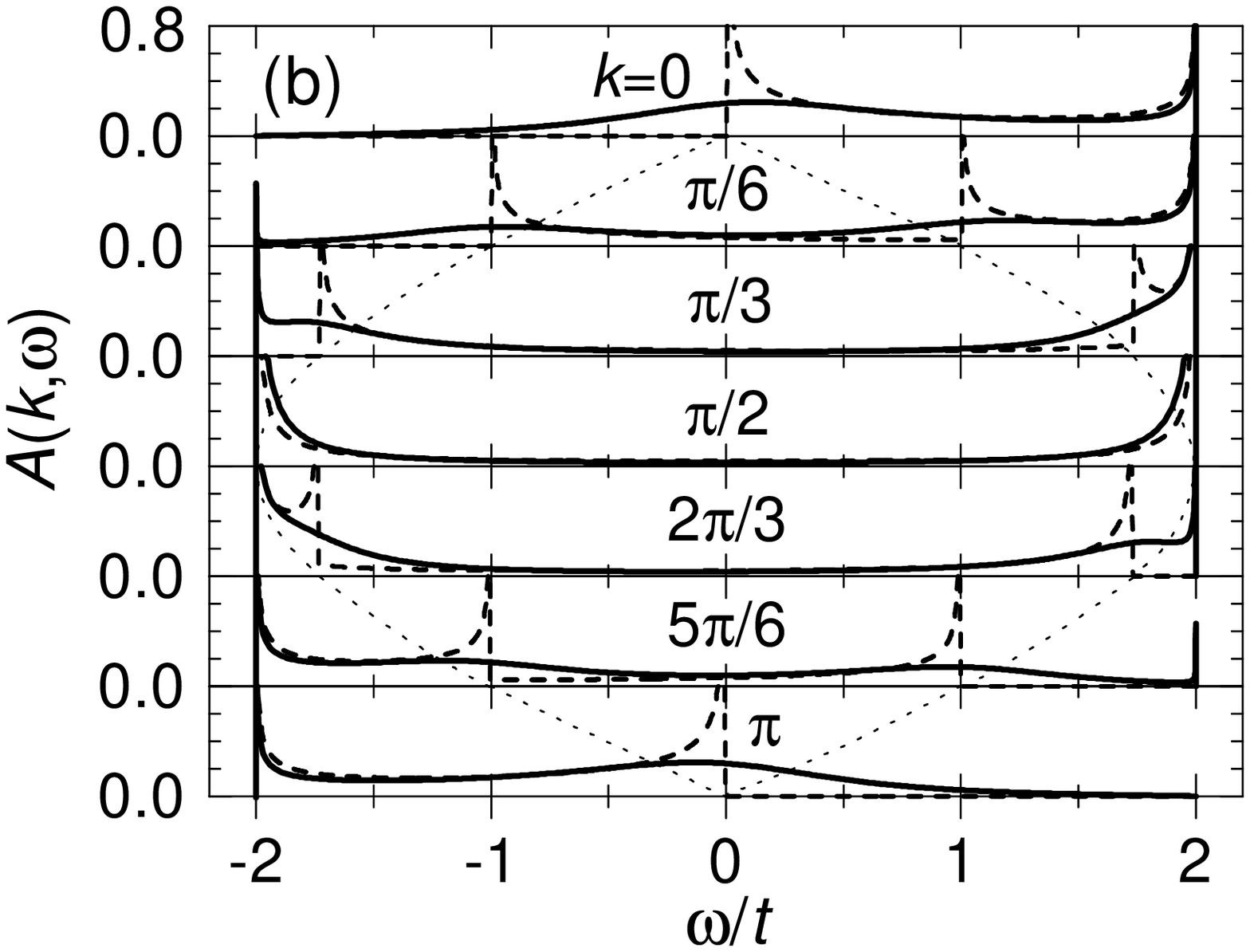,width=8cm,clip=}
\caption{(a) The weight $Z(Q)$ defined in eq.~(5).   The broken line is a fitted curve of the data for finite-size systems up to 24 sites (circle) at $J'$/$J$=0.  The solid line for $J'$/$J$=1/2 is obtained from eq.~(11).  (b) $A(k,\omega)$ for $J$/$t$$\rightarrow$0 obtained from eq.~(4).  The solid and broken lines are for $J'$/$J$=1/2 and 0, respectively.  The dotted lines denote holon dispersions.}
\label{fig1}
\end{center}
\end{figure}

The spectral functions obtained from eq.~(4) are shown in Fig.~1(b).  In the case of $J'$/$J$=0, the branch cut at $\omega$/$t$=$\pm$2$\sin k$ forms the holon dispersion.  The spinon dispersion is seen at $\omega$/$t$=2 with zero width at $k<\pi$/2.  In contrast, we find for $J'$/$J$=1/2 that (i) the dispersion with zero width extends to the region of $k>\pi$/2, and (ii) the singularity of the holon band is strongly suppressed.  The former (i) stems from the finite weight of $Z(Q)$ at $Q>\pi$/2, and the latter (ii) is related to the suppression of the divergence of $Z(Q)$ at $Q$=$\pi$/2.

\begin{figure}[t]
\begin{center}
\epsfig{file=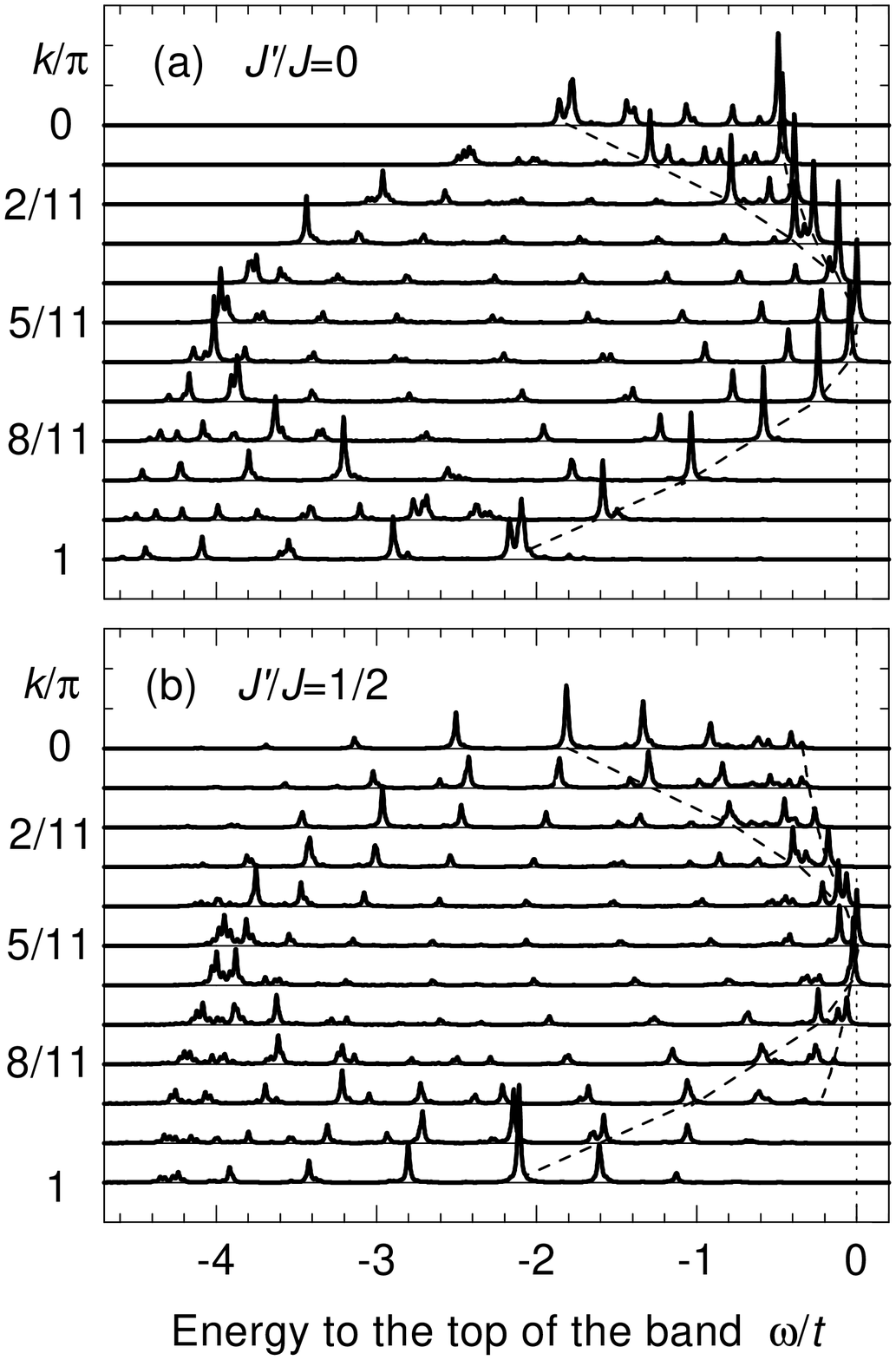,width=8cm,clip=}
\caption{$A(k,\omega)$ for a 22-site $t$-$J$-$J'$ ring with $J$/$t$=0.4. (a) $J'$/$J$=0 and (b) $J'$/$J$=1/2.  The dotted lines are guide to eyes.}
\label{fig2}
\end{center}
\end{figure}

To investigate $A(k,w)$ for finite values of $J$/$t$, we employ the exact diagonalization method for small clusters.  The results for a 22-site ring with $J$/$t$=0.4 are shown in Fig.~2.  The $\delta$ functions are convoluted with a Lorentzian broadening of 0.01$t$.  In Fig.~2(a), a dispersion with the width of about $\pi J$/2 is seen at low-energy region for $k<\pi$/2.  This is the spinon band.  The holon band is also clearly seen with the width of about 2$t$ and with large spectral weight. In contrast, at $J'$/$J$=1/2 (Fig.~2(b)), new structures emerge in the region of low-binding energy above the holon band at $k>\pi$/2, resulting in a symmetric behavior around $k$=$\pi$/2.  In addition, the spectral weight of the holon band is reduced.  These features are consistent with results for $J$/$t$$\rightarrow$0 limit.  The suppression of the holon weight was also discussed by Voit~\cite{Voit} for the Luther-Emery phase in the Luttinger liquid.  Therefore, this is a general feature in the spin-gapped system.  We also see the suppression of the weight at low-energy region of $k<\pi$/2 as compared to the case of $J'$/$J$=0.  $J'$ introduces frustrations that damp the spin excitations.  As a result, the spectral weight of the spinon branch is suppressed.  This effect is also seen in two-dimensional system~\cite{Shibata}.

Finally, we examine the quasi-particle weight of the hole defined as  $Z_{\rm h}$=$\left|\left<\psi_{L,0}^h \left| c_{k\uparrow} \right|\psi_L^0 \right>\right|^2$.  In the limit of $J$/$t$$\rightarrow$0, $Z_{\rm h}$=$Z$($\pi$/2)$\propto$$L^{-1}$$\rightarrow$0 for $J'$/$J$=1/2 as $L$$\rightarrow$$\infty$.  This means that there is no quasi-particle, and spin and charge are separated.  With increasing $J$/$t$, keeping $J'$/$J$=1/2, $Z_{\rm h}$ becomes finite as shown in Fig.~3.  This means the presence of a quasi-particle.  The study of hole-spin correlation indicates the formation of a bound state of the doped hole and an unpaired spin~\cite{Tohyama}.  The bound state carries both spin and charge degrees of freedom.  The same situation can be seen in ladder system where spin gap opens~\cite{Troyer}.  At larger value of $J$/$t$ ($>$4), $Z_{\rm h}$ decreases and becomes zero again at very large $J$/$t$.  This is caused by the localization of the hole which induces the separation of the hole and the unpaired spin~\cite{Tohyama,Takano}.

\begin{figure}[t]
\begin{center}
\epsfig{file=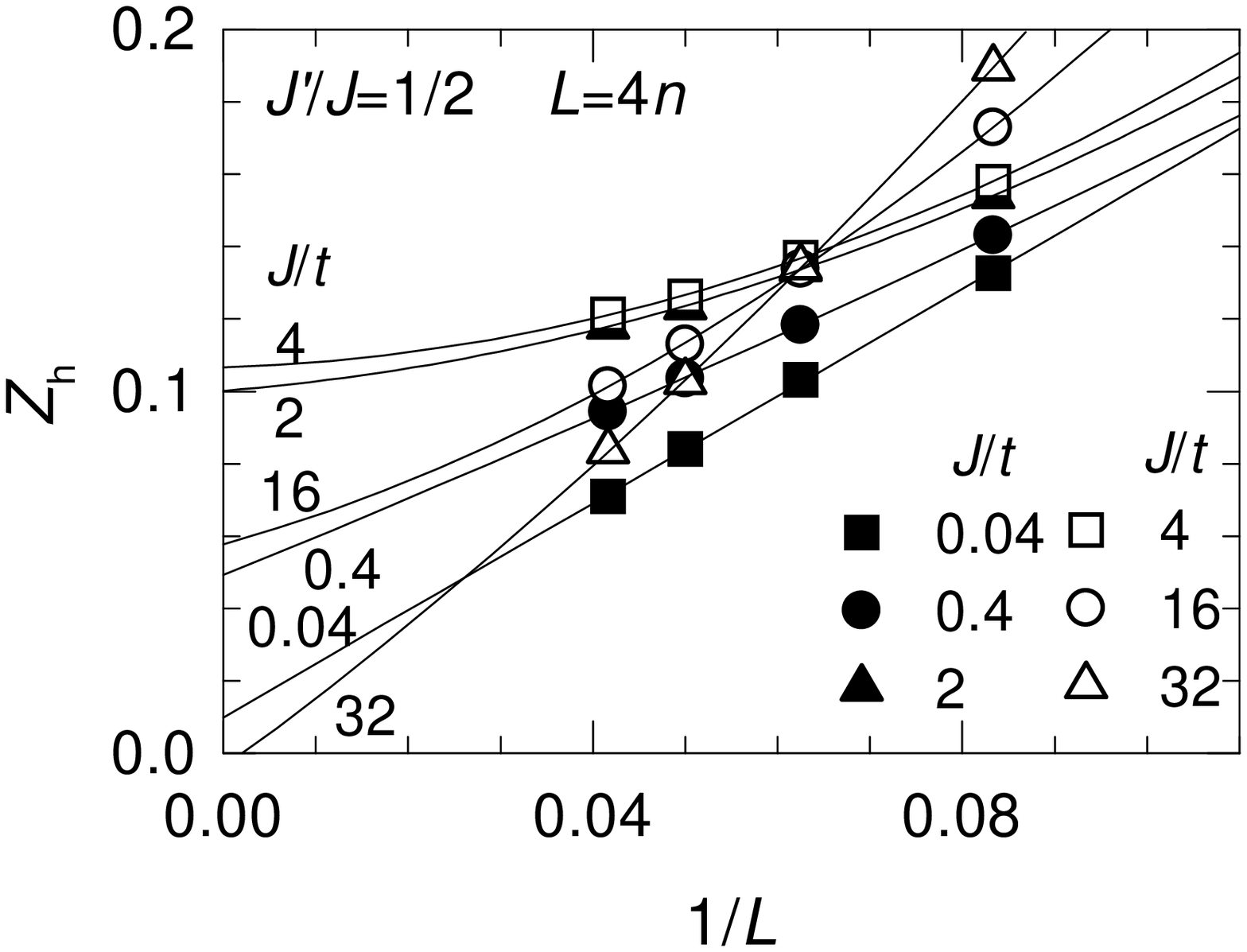,width=8cm,clip=}
\caption{System size dependence of quasi-particle weight $Z_{\rm h}$ at $J'$/$J$=1/2 up to $L$=24 ($L$=4$n$, $n$ being integer).  The data are fitted by a polynomial (solid lines).}
\label{fig3}
\end{center}
\end{figure}

In summary, we have investigate the single-hole excitation spectrum in the 1D $t$-$J$-$J'$ model with spin gap.  The presence of the spin gap has a strong effect on the spectrum.  The ARPES experiments are desired for 1D spin-gapped compounds such as CuGeO$_3$ and $\alpha$-NaV$_2$O$_5$ to confirm the effect.

The author (T.T.) would like to thank K. Sano for useful discussions.  This work was supported by a Grant-in-Aid for Scientific Research on Priority-Areas from the Ministry of Education, Science and Culture of Japan, and by the New Energy and Industrial Technology Development Organization (NEDO).  The numerical calculation were performed in the Supercomputer Center, ISSP, University of Tokyo, and the supercomputing facilities in IMR, Tohoku University.


\begin{thebibliography}{99}
\bibitem{Momentum}The ground state with the total momentum of zero ($K=0$) is degenerate with the lowest-energy state with $K$=$\pi$, namely, the anti-bonding state of eq.~(8).  $A(k,\omega)$ does not change even if we take the $K$=$\pi$ state.
\end{thebibliography}
\end{document}